# Dependency of Sliding Friction for Two Dimensional Systems on Electronegativity

Jianjun Wang[a,b,d], Avinash Tiwari[d], Yang Huang[b], Yu Jia[b,c,*], B.N.J. Persson[d]

[a] College of Science, Zhongyuan University of Technology, Zhengzhou, Henan 450007, China

[b] Key Laboratory for Special Functional Materials of Ministry of Education, and School of Materials Science and Engineering, Henan University, Kaifeng, Henan 475004, China

[c] International Laboratory for Quantum Functional Materials of Henan, and School of Physics, Zhengzhou University, Zhengzhou, Henan 450001, China

[d] PGI-1, FZ-Juelich, Germany, EU

**Abstract**

We study the role of electronegativity in sliding friction for five different two dimensional (2D) monolayer systems using density functional theory (DFT) with van der Waals (vdW) corrections. We show that the friction between the commensurate 2D layered systems depends strongly on the electronegativity difference of the involved atoms. All the 2D layered structures exhibit almost the same magnitude of friction force when sliding along the nonpolar path, independent of the material and the surface structures. In contrast, for sliding friction along the polar path, the friction force obeys a universal linear scaling law as a function of the electronegativity difference of its constituent atoms. Further analyses demonstrate that atomic dipoles in the 2D monolayers induced by the electronegativity difference enhance the corrugation of charge distribution and increase sliding barrier accordingly. Our studies reveal that electronegativity plays an important role in friction of low dimensional systems, and will provide a strategy for designing nanoscale devices further.

[*] jiayu@zzu.edu.cn (Y Jia)

# 1. Introduction

Understanding the origin of atom-scale friction processes and controlling them for designing nanotechnological devices pose a major challenge to physicists and nanotribologists alike due to the complex energy dissipation mechanisms and intricate interfacial interaction [1-3]. Two-dimensional (2D) layered materials, such as graphene, hexagonal boron nitride (*h*-BN), and hexagonal molybdenum disulfide ($MoS_2$), due to their strong intralayer chemical bonding compared with the weak interlayer physical adsorption interaction, can serve as solid lubricants to minimize friction and wear in high local pressures and boundary contact regime in a number of applications [4-6]. Moreover, 2D layered structures often have novel electronic properties widely researched in literature, which can further our understanding of their tribological behavior [7-11].

Friction is ultimately governed by the atomistic interactions dominated by quantum mechanics [2, 3, 12]. Several studies have demonstrated that electronic structure, charge distribution, and even spin degree of freedom can influence the friction behaviors of low dimensional system [6, 11-13]. Electronegativity plays a significant role in the electronic structure, and electronegativity difference among constituent atoms has an important influence on the physical and chemical properties of materials [14]. For example, although h-BN and graphene are isoelectronic and isostructural, the former is an insulator with a wide band gap of around 6 eV, while the latter is a zero band gap semimetal [15]. This is because the large electronegativity difference between B and N atoms displaces a shared pair of electrons towards the N atom. Thus, electronegativity has an influence on the structure and hence also on the friction. The relationship between electronegativity and friction in low dimensional system was studied in Ref. [16-21]. Experiments have shown that the sliding friction in insulating multiwall BN nanotubes (BNNTs) is orders of magnitude stronger than that of semimetallic C nanotubes (CNTs), which were attributed to increased potential barrier caused by the charge localization induced by electronegativity difference between the B and N atoms in BNNTs [16]. This localization effect increases the corrugation amplitude of the interfacial potential. Ab initio molecular dynamics calculation showed that the coefficient of friction (COF) of liquid water sliding on h-BN is about three times larger than that of on graphene, which was ascribed to the greater corrugation of the energy landscape of h-BN arising from specific electronic structure effect [17]. From DFT calculations it was found that the vdW interaction determines the interlayer binding and the electrostatic interaction mainly influences the sliding barrier of bilayer h-BN [18-20]. They speculated that a highly anisotropic interfacial friction should exist for the h-BN bilayer [19]. DFT calculations found that constraints on atomic motion can be employed to tune the contribution of electrostatic interactions and dispersive forces to the sliding energy profile, ultimately leading to different sliding pathways in bilayers of graphene and h-BN [20]. All the above results confirm that the polarity plays an important role in friction of low dimensional systems. However, no detailed investigations of the connection between polarity and friction properties have been published.

In this research work, we systematically investigate the role of electronegativity on friction in 2D layered systems. We show that when two monolayers slide relative to each another, the atom electronegativity difference along the sliding path strongly influence the interfacial friction properties. We show that the friction scales linearly with electronegativity difference in the polar sliding path.

## 2 Methodology

Vienna Ab-initio Simulation Package (VASP) code based on the projector augmented-wave (PAW) method was employed in the calculations [22–24]. The exchange-correlation interactions were treated with the generalized gradient approximation (GGA) of Perdew, Burke, and Ernzerhof (PBE) [25], with a vdW correction determined by the many-body dispersion (MBD) method [26]. An energy cut-off of 600 eV and 21×21×1 Monkhorst-Pack grids were selected for 2D irreducible Brillouin-Zone integration [27]. The convergence thresholds for total energy and Hellmann-Feynman forces are $10^{-5}$ eV and 0.01 eV/Å, respectively. A vacuum space of at least 20 Å was set to avoid the intercell interactions.

Based on the Prandtl-Tomlinson model and its extensions, several researchers have developed methods to compute friction using DFT calculations. In these methods, typical friction parameters, such as shear force, COF and potential energy surface (PES), can be obtained by calculating the energy barrier or sliding energy corrugation along the sliding path [6, 11, 12, 18, 28, 29]. It should be noted that these methods only evaluate the maximum energy barrier along the sliding path, but did not consider the energy dissipation while sliding. Therefore, the friction calculated by these methods is static friction or break-loose force. In our work, the COF and PES are separately calculated by the above methods [28, 29]. However, if it is assumed that all the energy needed to reach the top of the energy barrier is dissipate before climbing the next barrier, and then present calculation also gives information about the kinetic friction [30].

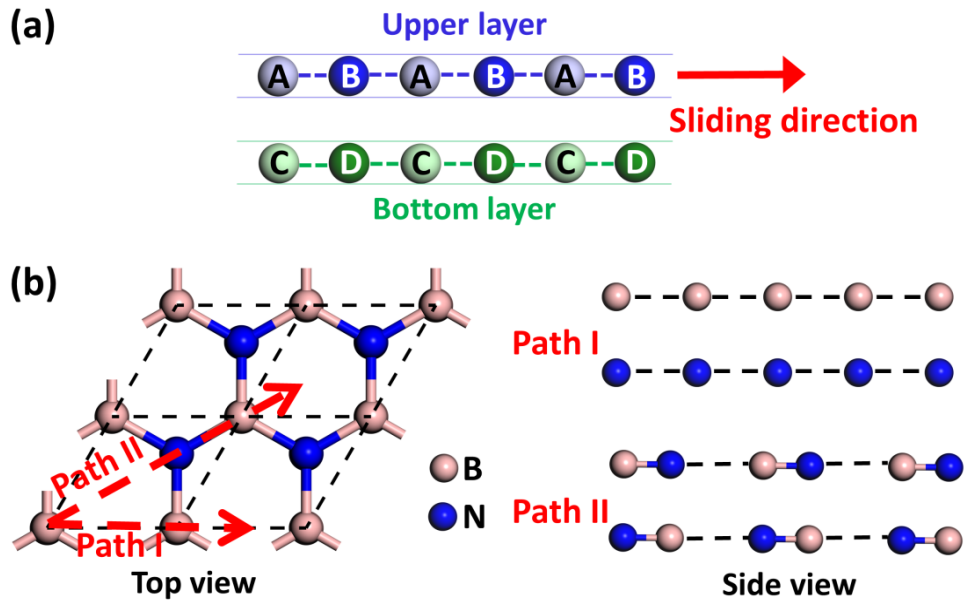

**Fig. 1.** Schematic diagram of polar and nonpolar friction paths. (a) General atomic arrangement of a sliding interface along the sliding direction. (b) Top and side views of two sheets of *h*-BN with AA' stacking. Path I and II are two highly symmetric paths along the basis vector and its diagonal direction, respectively.

## 3. Results and Discussion

In order to study the influence of polarity on friction, we first discuss the polarity character in the sliding path. Fig .1(a) shows a schematic diagram of the sliding interface between two monolayers. The atoms A, B form the upper layers, and C, D form the bottom layers arrange alternatively along the sliding direction. The total polarity of the sliding system along the sliding direction is then the sum of the polarity of each single layer in that direction, which is decided by the electronegativity differences between A and B atoms, and C and D atoms. Here, bilayer *h*-BN is chosen as an example to illustrate the polarity differences in different paths, as shown in Fig. 1(b). The most stable AA' stacking eclipsed with B over N atoms was chosen as an initial structure [18, 19, 31], and two highly symmetric directions were chosen as sliding paths. From the cross-section view, we can see that along path I, there only one kind of B (N) atom in the upper (bottom) layer, and so the polarity is zero along this direction. However, B and N atoms alternately arrange in upper and bottom layers along path II, which give rise to dipoles in both the upper and bottom layers. Therefore, we denote the path I and II as nonpolar and polar paths, respectively.

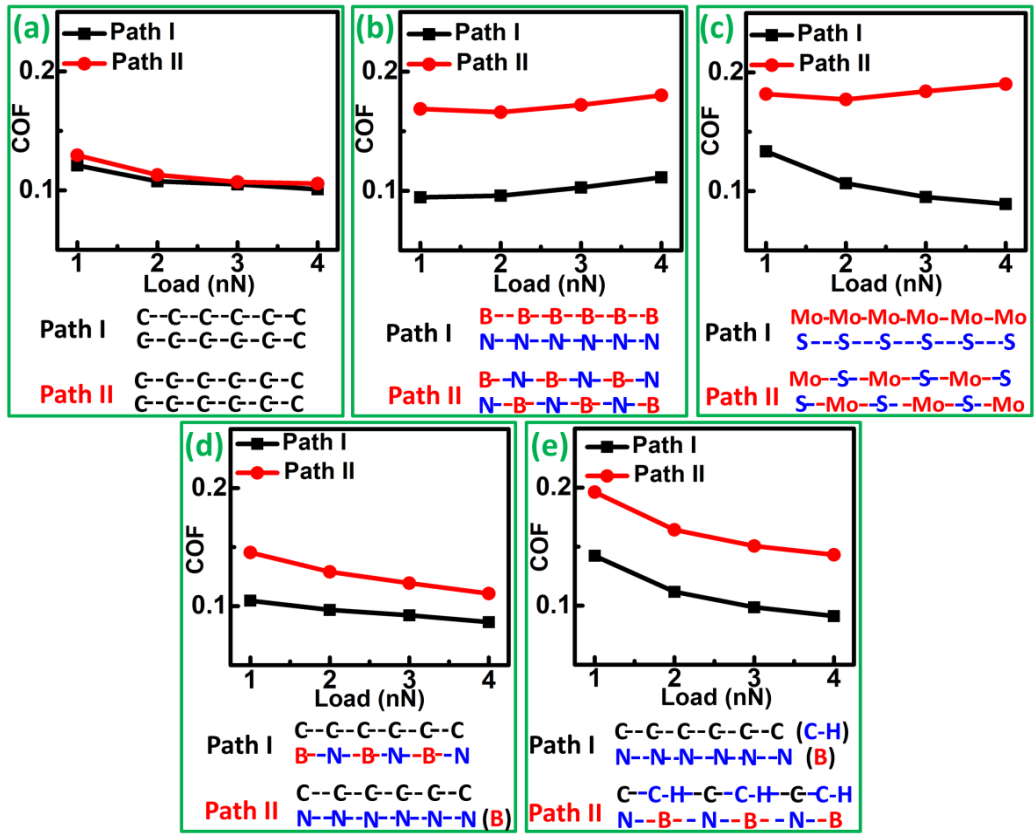

**Fig. 2.** Coefficient of friction (COF) as a function of normal load along polar and nonpolar paths. (a) graphene/graphene (b) $h$-BN/$h$-BN (c) $MoS_2$/$MoS_2$ (d) graphene/$h$-BN and (e) H-graphene/$h$-BN systems, respectively.

The optimized lattice parameters of graphene, $h$-BN, and $MoS_2$ are 2.47, 2.51, and 3.18 Å, respectively. Based on the lattice parameters, the stable bilayer stacking models for the three systems were obtained (see Table S1 of supporting materials), which is supported by other study [18, 19, 32]. Two sheets of the 2D monolayer were moved relative to each other along the path I and II to simulate the sliding process. The COF as a function of the load is shown in Fig. 2. For the identical pristine layers of graphene/graphene (Fig. 2(a)), $h$-BN/$h$-BN (Fig. 2(b)), and $MoS_2$/$MoS_2$ (Fig. 2(c)), the COF fall in a range of 0.07-0.2, which is in agreement with earlier studies [5, 11]. Among these three systems, the nonpolar graphene/graphene system exhibits isotropic friction behavior. However, in the polar $h$-BN and MoS$_2$ systems, the COF exhibit anisotropic behavior, in which the COF along polar path II is almost two times larger than for the nonpolar path. These results clearly show that the polarity plays an important role in the friction between two polar planes.

We next investigated the friction between polar and nonpolar sheets by using the isoelectronic interface of graphene on $h$-BN (graphene/$h$-BN). The computational model was simplified by enlarging the lattice constant of graphene to be equal to that of $h$-BN so that the influence of incommensuration and Moiré patterns on friction can be cancelled. It should be noticed that although the upper layer is nonpolar, path II is still polar due to dipoles that exist in the bottom layer along the sliding path (Fig. 2(d)). Similar to the polar/polar interface, the COF in the nonpolar/polar system is also anisotropic, with a larger COF in polar path. These results indicate that even if dipoles exist only in one layer of the sliding interface, the polarity still has a great influence on the interfacial friction.

Based on the graphene/h-BN system, we further consider the friction modulation through tuning polarity of graphene. For this purpose, we induced polarity into graphene lattice by single side half hydrogenation (H-graphene). The covalent C-C bond in graphene turns into a partially ionic bond in the single side half hydrogenated graphene (H-graphene), with the charge transfer of 0.35 e (0.21 e for Bader analysis). The COF between H-graphene and $h$-BN (H-graphene/$h$-BN) was calculated and is shown in Fig. 2(e). It is apparent that the COF along polar path II is larger than that of the nonpolar path I. More importantly, the H-graphene/$h$-BN system has a larger COF than that of the graphene/$h$-BN, which is consistent with the increase of the polarity induced by H passivation. The calculations provide new insight for understanding and tuning friction by surface modification of monolayers.

The variation of the interaction energy as a function of the relative lateral position of the two surfaces in contact can be represented by PES, which is relevant for the friction in the case when no external load is applied [29]. At the fundamental level, the corrugation of PES determines the intrinsic resistance to sliding and gives an indication of the maximum energy that can be dissipated during frictional processes. From the Fig. 3(a) and (b) it is observed that the *h*-BN system has larger PES corrugation than that of graphene system. To examine the effect of sliding direction on friction, we plotted the potential profiles along the two symmetric directions, indicated in the middle of the Fig. 3. In the graphene/graphene system, the two paths exhibit almost the same potential corrugation. However, for *h*-BN/*h*-BN system, the potential corrugation along path II is about two times larger than that of along path I, which result in different friction behaviors as observed earlier in Fig. 2. The lateral forces acting on the slider dragged along the two symmetric directions were also calculated and are shown at the bottom of Fig. 3. For the graphene system, the maximum friction force values are almost equal for the two different paths. On the contrary, the *h*-BN system exhibits larger friction force along path II than that of path I. These results further demonstrate the anisotropic friction behavior of the *h*-BN system, which corroborate results of COF in Fig. 2.

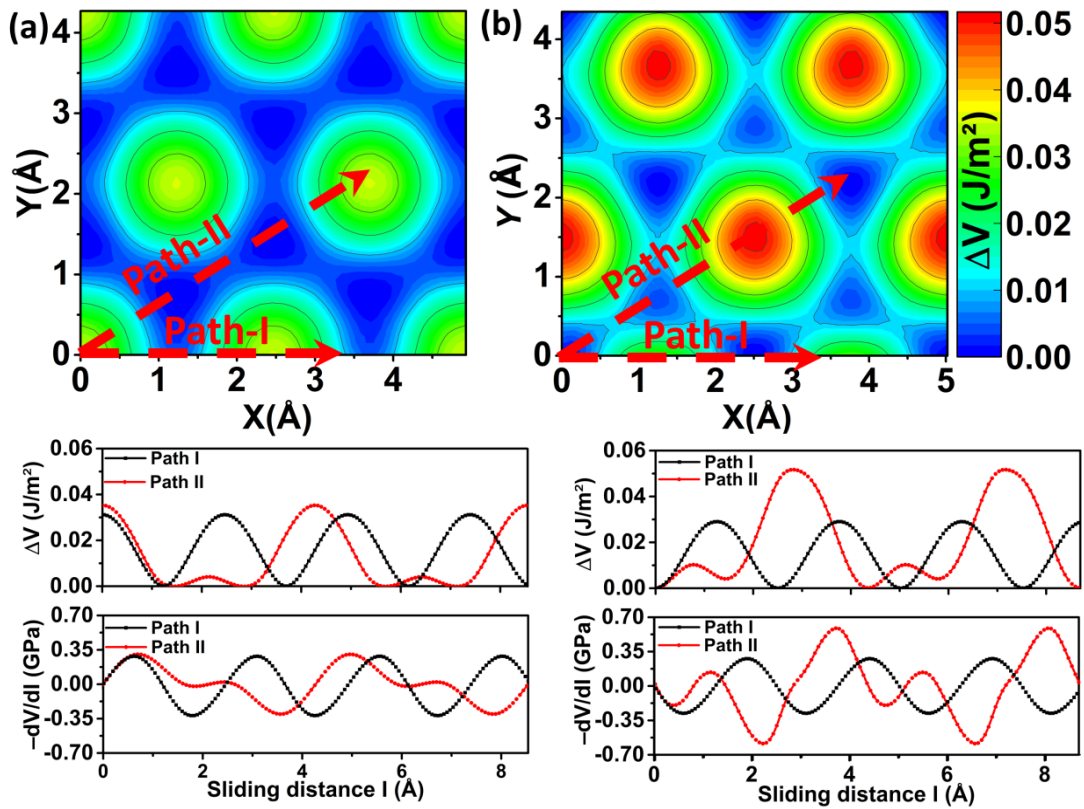

**Fig. 3.** (a) and (b) are the potential energy surfaces (PESs) in graphene/graphene and *h*-BN/*h*-BN systems respectively. The absolute PES minimum (in blue) is taken as a reference, and a common energy scale is used. The corresponding potential barriers and friction forces along the two chosen

paths are plotted under each PES.

From the above calculations, we conclude that the frictional properties exhibited by 2D systems both with and without external load, are dependent on the polarity and hence the directional sliding. Note that, the sliding potential barrier $\Delta V = \Delta E + F_N \Delta h$, where $F_N$ is the external load, $\Delta E$ and $\Delta h$ are the changes in the binding energy and the interlayer separation during sliding, depend on (a) the variation of the adsorption bond energy, and (b) the work done to overcome interlayer separation during sliding under the external load $F_N$. When no external load is applied, the friction is decided by the $\Delta E$. However, for the normal loads used in our study, the major contribution to the sliding barrier comes from the work done against the external force, and the change in $\Delta E$ accounts for only a small part (see fig. S1).

To understand the origin of the anisotropic friction behavior of the *h*-BN polar system, we calculated the charge density difference of *h*-BN system by the formula $\Delta \rho = \rho_{BN} - \rho_B - \rho_N$, where $\rho_{BN}$, $\rho_B$ and $\rho_N$ are the charge densities of *h*-BN, free B and N atoms, respectively. Fig. 4(a) shows that charge accumulates around N atoms from neighbor B atoms. The Mulliken population analysis estimates the transferred charge to be 0.84 e (2.18 e for Bader analysis). We further calculated the interlayer charge density difference of the stable bilayer *h*-BN. From the Fig. 4(b) we can see that when two sheets of *h*-BN approach each other, the intralayer charge transfer from B to N atoms is further enhanced. However, this effect is noticeably different in the two paths as seen in cross-sectional views of the charge density difference. Along the nonpolar path I, the net positive charge on the upper layer and the negative charge in the bottom layer separately distribute along the sliding direction, which corresponds to a small charge corrugation and smooth sliding barrier. However, along the polar path II (Fig. 4(d)), electron-deficient B and electron-rich N atoms alternately appear in both bottom and upper layers, which causes a large fluctuation of charge distribution along this direction. Of course, the larger the electronegativity difference, the larger corrugation of charge distribution and friction. In contrast, there are only C atoms in graphene system and all sliding paths are nonpolar causing the graphene to manifest friction isotropy.

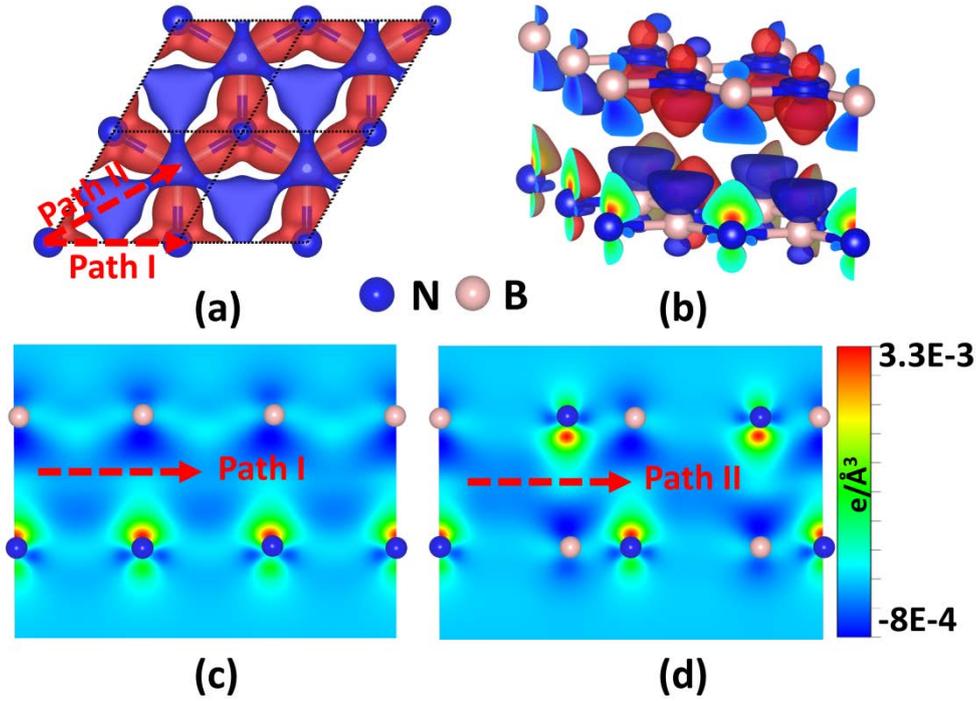

**Fig. 4.** The charge density difference of (a) single layer and (b) bilayer of *h*-BN, with pink and blue indicating charge accumulation and depletion regions. The isosurface charge density is taken as 0.01 electrons/Å$^3$. (c) And (d) are the corresponding cross-sectional views of the bilayer system along path I and path II, respectively.

To establish a clear relationship between electronegativity difference and interfacial friction, we further quantitatively compared the friction properties for all calculated systems, as shown in Fig. 5. The comparisons of COF under the load of 3 nN/unit-cell in different systems exhibit two characters. One character is that along the nonpolar path I, the COFs are almost equal only with a slight fluctuation. Interesting is to note that the COF of about ~0.1 belongs to the scale of typical vdW friction [6, 33, 34], indicating that the vdW interaction is the main cause of friction along the nonpolar path I. These values are consistent and are corroborated by reported values of vdW interaction in earlier research studies [18, 19, 33]. The other character is that the polar path exhibits larger friction than that of nonpolar path for all systems.

To better understand the different COF along the polar direction II, we further extracted the functional relation between charge transfers and friction, as shown in Fig. 5 (b). It should be pointed out here that the charge transfer is the sum of the intralayer charge transfer in the upper and lower surfaces. As the different charge analysis method may yields some deviation, the charges transfers were separately calculated by both Mulliken population and Bader charge analysis. Although the two methods provide different amount of charge transfers, the change of the amount of charge transfer with respect to the friction force is quite similar, with errors falling within a small linear window. It is interesting that a linear functional relationship has been found between averaged frictional forces and the amount of intralayer charges transfer. It should be note that, although the COF in polar direction is larger for $MoS_2$, the shear force per unit cell is small. This because the unit of the normal load is nN/unit-cell and the $MoS_2$ has large cell. Compared to the nonpolar path I, the added friction in the polar path II can now be safely attributed to the enhanced charge density corrugation that appear in the polar path II.

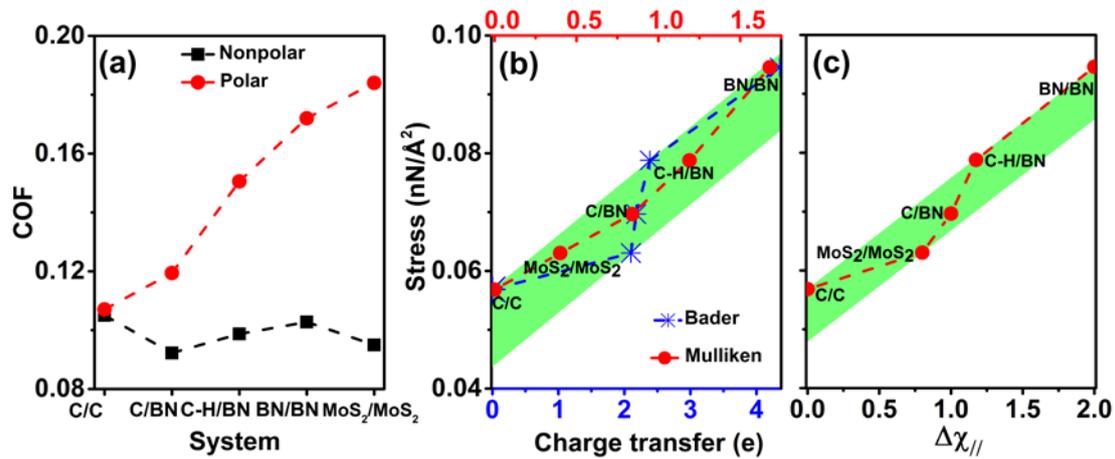

**Fig. 5.** (a) Comparisons of COFs under the external load of 3 nN/unit-cell among different 2D bilayer systems. The shear stress as a function of (b) charge transfer and (c) electronegativity difference $\Delta\chi_{//}$.

The relationship between the electronegative difference and friction is further discussed as follows. To establish a functional relation, we first quantize the electronegativity difference of a 2D interface along a sliding direction. We define the total electronegativity difference of the system along a path as $\Delta\chi_{//} = |\chi_A - \chi_B| + |\chi_C - \chi_D|$, where $\chi_X$ represents electronegativity of the X atom. By the definition, the greater the electronegativity differences between its constituent atoms of each layer, the greater $\Delta\chi_{//}$. Similar to charge transfer in Fig. 5(b), the frictional forces increase with the increasing of $\Delta\chi_{//}$, which obeys a linear scaling law (Fig. 5(c)).

From the above analysis, one can see that the electronegativity difference in one preferred direction dominates the friction behavior along the sliding path for 2D systems. We denote this friction as electronegativity induced friction, which obeys a linear scaling law. It should be emphasized here that, all of the sliding interfaces in our calculation are commensurate, and the obtained scaling law are only suited for the commensurate sliding. In the incommensurate interface, such as identical double layers with relative rotation, the polarity effect will be cancelled, and the ultra-low friction will be observed.

## 4 Conclusions

Using the DFT including dispersion correction, the role of electronegativity in interfacial friction has been investigated. Five systems with different polarities (Graphene/Graphene, Graphene/$h$-BN, $h$-BN/$h$-BN, H-Graphite/$h$-BN and $MoS_2$/$MoS_2$) were chosen in the study. Our results show that all of the systems exhibit almost same friction along nonpolar sliding path I, which is consistent with the similar values of vdW interactions. In contrast, the interfacial friction force along polar path obeys a linear scaling law with the electronegativity difference among its constituent atoms. This linear scaling relationship between friction and electronegativity difference can be extended to other low-dimensional materials such as multiwall nanotubes and nanosheets, which have important applications in nanotechnology. The results of this paper provide new ideas for the design of nanoscale devices. For example, one can provide specific sliding channels through the design of polar or nonpolar paths in low-dimensional materials. Therefore, the research has important scientific significance and application value.

**Acknowledgements**

The work was supported by the National Natural Science Foundation of China (Grant No. U1604131, 11774078, 11805295).

**References**

[1] B. N. J. Persson, Sliding Friction: Physical Principles and Applications (Springer, Heidelberg, 2000).

[2] M. Urbakh and E. Meyer, Nat. Mater. **9**, 8–10 (2010).

[3] O. M. Braun and A. G. Naumovets, Surf. Sci. Rep. **60**, 79–158 (2006).

[4] C. Lee, Q. Li, W. Kalb, X. Liu, H. Berger, R. W. Carpick and J. Hone, Science **328**, 76 (2010).

[5] J. C. Spear, B. W. Ewers and J. D. Batteas, Nano Today **10**, 301–314 (2015).

[6] S. Cahangirov, C. Ataca, M. Topsakal, H. Sahin and S. Ciraci. Phys. Rev. Lett. **108**, 126103 (2012).

[7] X. Feng, S. Kwon, J. Y. Park and M. Salmeron, ACS Nano **7**, 1718–1724 (2013).

[8] O. Hod, E. Meyer, Q. Zheng and M. Urbakh, Nature **563**, 485–492 (2018).

[9] S. Kwon, J. H. Ko, K. J. Jeon, Y. H. Kim and J. Y. Park, Nano Lett. **12**, 6043–6048 (2012).

[10] R. C. Sinclair, J. L. Suter and P. V. Coveney, Adv. Mater. **30**, 1–7 (2018).

[11] J. Wang, J. Li, L. Fang, Q. Sun and Y. Jia, Tribol. Lett. **55**, 405 (2014).

[12] M. Wolloch, G. Levita, P. Restuccia and M. C. Righi, Phys. Rev. Lett. **121**, 026804 (2018).

[13] B. Wolter, Y. Yoshida, Kubetzka A, S.–W. Hla, K. Bergmann and R. Wiesendanger, Phys. Rev. Lett. **109**, 116102 (2012).

[14] T. Sanderson, J. Am. Chem. Soc. **105**, 2259–2261 (1983).

[15] D. Golberg, Y. Bando, Y. Huang, T. Terao, M. Mitome, C .Tang and C. Zhi, ACS Nano **4**, 2979–2993 (2010).

[16] A. Niguès, A. Siria, P. Vincent, P. Poncharal and L. Bocquet, Nat. Mater. **13**, 688–693 (2014).

[17] G. Tocci, L. Joly and A. Michaelides, Nano Lett. **14**, 6872–6877 (2014).

[18] N. Marom, J. Bernstein, J. Garel, A. Tkatchenko, E. Joselevich, L. Kronik and O. Hod, Phys. Rev. Lett. **105**, 046801 (2010).

[19] O. Hod, J. Chem. Theory Comput. **8**, 1360–1369 (2012).

[20] W. Gao and A. Tkatchenko, Phys. Rev. Lett. **114**, 096101 (2015).

[21] A. Falin, Q. Cai, E. J. G. Santos, D. Scullion, D. Qian, R. Zhang, Z. Yang, S. Huang, K. Watanabe, T. Taniguchi, M. R. Barnett, Y. Chen, R. S. Ruoff and L. H. Li, Nat. Commun. **8**, 15815 (2017).

[22] G. Kress and J. Furthmüller, Phys. Rev. B **54**, 11169 (1996).

[23] G. Kresse and D. Joubert, Phys. Rev. B **59**, 1758 (1998).

[24] P. E. Blöchl, Phys. Rev. B **50**, 17953 (1994).

[25] J. P. Perdew, K. Burke and M. Ernzerhof, Phys. Rev. Lett. **77**, 3865 (1996).

[26] A. Tkatchenko, DiStasio R. A. Jr, R. Car and M. Scheffler, Phys. Rev. Lett. **108**, 236402 (2012).

[27] H. J. Monkhorst and J. D. Pack, Phys. Rev. B **13**, 5188 (1976).

[28] W. Zhong and D. Tománek, Phys. Rev. Lett. **64**, 3054 (1990).

[29] G. Zilibotti and M. C. Righi, Langmuir **27**, 6862–6867 (2011).

[30] B. N. J. Persson, Tribol. Lett. **68**, 28 (2020).

[31] G. Constantinescu, A. Kuc and T. Heine, Phys. Rev. Lett. **111**, 036104 (2013).

[32] J. He, K. Hummer and C. Franchini, Phys. Rev. B **89**, 075409 (2014).

[33] S. Zhou, J. Han, S. Dai, J. Sun and D. J. Srolovitz, Phys. Rev. B **92**, 155438 (2015).

[34] B. Li, J. Yin, X. Liu, H. Wu, J. Li, X. Li and W. Guo, Nat. Nanotech. **14**, 567–572 (2019).

# Dependency of Sliding Friction for Two Dimensional Systems on Electronegativity

Jianjun Wang[a,b,d], Avinash Tiwari[d], Yang Huang[b], Yu Jia[b, c,*], Bo N. J. Persson[d]

[a] *College of Science, Zhongyuan University of Technology, Zhengzhou, Henan 450007, China*

[b] *Key Laboratory for Special Functional Materials of Ministry of Education, and School of Materials Science and Engineering, Henan University, Kaifeng, Henan 475004, China*

[c] *International Laboratory for Quantum Functional Materials of Henan, and School of Physics, Zhengzhou University, Zhengzhou, Henan 450001, China*

[d] *PGI-1, FZ-Juelich, Germany, EU*

---

[*] jiayu@zzu.edu.cn (Y Jia)

**Table S1**

Table S1. The structure parameters of stable bilayer stacking

|  | stable stacking | binding energy (meV/unit-cell) | separation space (Å) |
|---|---|---|---|
| Graphene/graphene | AB | 92.77 | 3.36 |
| $h$-BN/$h$-BN | AA' | 109.08 | 3.33 |
| $MoS_2$/$MoS_2$ | AA' | 150.16 | 3.13 |

**Figure S1**

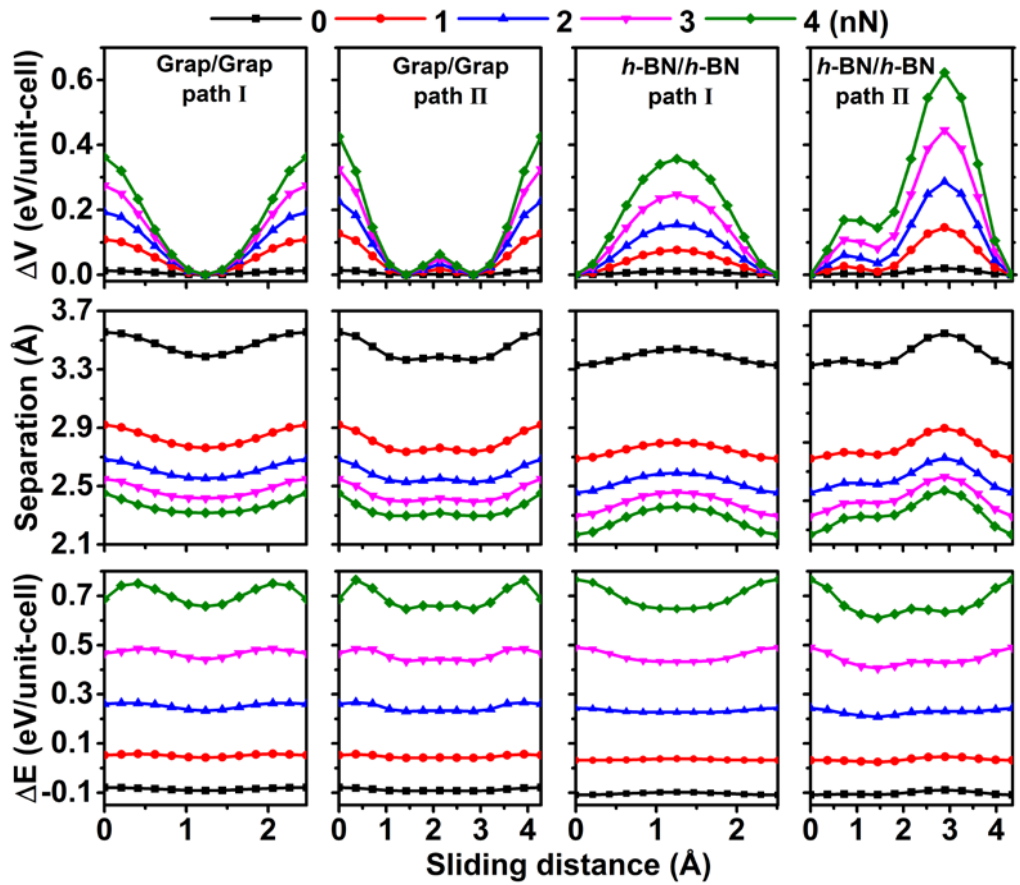

Figure S1. Sliding barrier ΔV, interlayer separation and binding energy ΔE as a function of sliding distance for graphene/graphene and $h$-BN/$h$-BN systems along path I and II, respectively.